\documentclass{iau} 
\usepackage{graphicx}

\title[Optical light curves of FUor and FUor-like objects] 
{Optical light curves of FUor and FUor-like objects}

\author[Evgeni Semkov, Stoyanka Peneva \& Sunay Ibryamov]   
{Evgeni Semkov$^1$, Stoyanka Peneva$^1$
 \and Sunay Ibryamov$^1$$^,$$^2$}

\affiliation{$^1$Institute of Astronomy and National Astronomical Observatory, Bulgarian Academy of Sciences, Sofia, Bulgaria 
\\ email: {\tt esemkov@astro.bas.bg} \\[\affilskip]
$^2$Department of Theoretical and Applied Physics, University of Shumen, Shumen, Bulgaria}

\pubyear{2017}
\volume{325}  
\jname{Astroinformatics 2016}
\editors{M. Brescia, G. Djorgovski, E. Feigelson, G. Longo \& S. Cavuoti, eds.}
\begin{document}

\maketitle

\begin{abstract}
Using recent data from photometric monitoring and data from the photographic plate archives we aim to study, the long-term photometric behavior of FUors. 
The construction of the historical light curves of FUors could be very important for determining the beginning of the outburst, the time to reach the maximum light, the rate of increase and decrease in brightness, the pre-outburst variability of the star. 
Our CCD photometric observations were performed with the telescopes of the Rozhen (Bulgaria) and  Skinakas (Crete, Greece) observatories. Most suitable for long-term photometric study are the plate archives of the big Schmidt telescopes, as the telescopes at Kiso Observatory, Asiago Observatory, Palomar Observatory and others.
In comparing our results with light curves of the well-studied FUors, we conclude that every new FUor object shows different photometric behavior. Each known FUor has a different rate of increase and decrease in brightness and a different light curve shape.

\end{abstract}

\firstsection 
\section{Introduction}

The young eruptive objects such as FU Orionis (here after FUor) are very rare, but with a significant role in the stellar evolution.
All known FUors share the same defining characteristics: a $\Delta$$V$$\approx$4-6 magnitudes outburst amplitude, association with reflection nebulae, location in regions of active star formation, an F-G supergiant spectrum during the outburst (\cite[Audard et al. 2014]{Audar_etal14};
\cite[Reipurth \& Aspin 2010]{ReiAs10}).
FUor stars seem to be related to the low-mass pre-main sequence objects (T Tauri stars), which have massive circumstellar disks. 
The widespread explanation of the FUor phenomenon is a sizable increase in accretion rate from the circumstellar disc onto the stellar surface.
The cause of increased accretion is still being discussed.
But the possible triggering mechanisms of FUor outburst could be: thermal or gravitational instability in the circumstellar disk 
(\cite[Hartmann \& Kenyon 1996]{Hart1966})  and the interactions of the circumstellar disk with a giant planet or nearby stellar companion on an eccentric orbit 
(\cite[Lodato \& Clarke 2004]{Lod2004}; \cite [Pfalzner 2008]{Pfa2008}).

The construction of the historical light curves of FUors could be very important to study the photometric evolution of the objects, for determining the exact moment of the beginning of the outburst, the time to reach the maximum light and the time spent in maximum light.
Another important option is to study the pre-outburst variability of the FUor objects.

\section{Observations}

Our CCD photometric observations of FUor objects were performed with the 2 m RCC, the 50/70 cm Schmidt, and the 60 cm Cassegrain telescopes of the National Astronomical Observatory Rozhen (Bulgaria) and with the 1.3 m RC telescope of the Skinakas Observatory\footnote{Skinakas Observatory is a collaborative project of the University of Crete, the Foundation for Research and Technology - Hellas, and the Max-Planck-Institut f\"{u}r Extraterrestrische Physik.} of the Institute of Astronomy, University of Crete (Greece).  
The technical parameters for the CCD cameras used, observational procedure and data reduction process are described in \cite[Ibryamov et al. (2015)]{Ibryamov2015}.

The only possibility for long-time photometric study is a search in the photographic plate archives at the astronomical observatories around the world.
Most suitable for this purpose are the plate archives of the big Schmidt telescopes that have a large field of view.
In this paper we present photometric data obtained from the photographic plate archives of the 105/150 cm Schmidt telescope at the Kiso Observatory (Japan) and the 67/92 cm Schmidt telescope at the Asiago Observatory (Italy).
We also used the digitized plates from the Palomar Schmidt telescope, available via the website of the Space Telescope Science Institute.

\section{Results and discussion}

In this section we present results from the long-term photometric study of three FUor objects.
All three objects were discovered recently and the available data for their photometric behavior are still incomplete.

\subsection{V2493 Cyg}

The outburst of V2493 Cyg was discovered during the summer of 2010 (\cite[Semkov et al. 2010]{Semkov2010}; \cite[Miller et al. 2011]{Miller2011}) in the dark clouds  between NGC 7000 and IC 5070 (so-called "Gulf of Mexico").
Subsequent photometric and spectral observations (\cite[K{\'o}sp{\'a}l et al. 2011]{Kospal2011}; \cite[Semkov et al. 2012]{semkov2012}; \cite[Baek et al. 2015]{Baek2015}) indicate that the object can be definitely assigned to the class of FUors.
The $BVRI$ light curves of V2493 Cyg from the collected photometric data are plotted in Fig. 1. 
The filled diamonds represent our CCD observations from Rozhen and Skinakas observatories,
the filled circles CCD observations from the 48 inch Samuel Oschin telescope at Palomar Observatory (\cite[Miller et al. 2011]{Miller2011}), 
the open diamonds photographic data from the Asiago Schmidt telescopes,
the open squares photographic data from the Palomar Schmidt telescope,
the filled squares photographic data from the Byurakan Schmidt telescope
and the open circles photographic data from the Rozhen 2-m RCC telescope.

\begin{figure}[b]
\begin{center}
 \includegraphics[width=13cm]{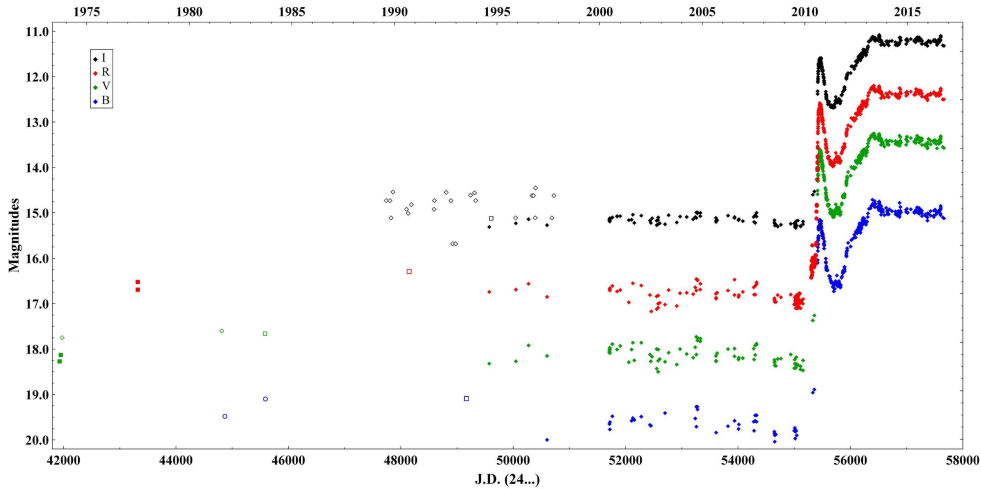} 
 \caption{Historical $BVRI$ light curves of V2493 Cyg for the period September 1973 - November 2016}
   \label{fig1}
\end{center}
\end{figure}

The photometric observation obtained before the outburst displayed only small amplitude variations in all pass-bands typical of T Tauri stars.  
Acoording to our data the outburst started sometime before May 2010, and reached the first maximum value at the period September - October 2010.  
Since October 2010, a slow fading was observed and up to May 2011 the star brightness decreased by 1.4 mag. ($V$). 
Since the autumn of 2011, another light increase occurred and the star became brighter by 1.8 mag. ($V$) until April 2013. 
From the spring of 2011 up to now the star keeps its maximum brightness showing a little bit fluctuations around it.
Therefore, we have observed a classical outburst from FUor type, which should continue over the next few decades.

\subsection{V582 Aur}

The discovery of V582 Aur was reported by the amateur astronomer Anton Khruslov. 
The star is located in a region of active star formation near Auriga OB2 association.
According to \cite[Samus (2009)]{Samus2009} the increase in brightness of the star started between 1982 and 1986. 
\cite[Munari et al. (2009)]{Munari2009} obtained the first spectrum of V582 Aur, which confirms the FUor nature of the star (presence of absorption lines of the Balmer series, Na I D and Ba II ($\lambda$ 6496)). 
On the basis of spectral and photometric (\cite[Semkov et al. 2013]{Semkov2013}) data we prove that the star is a FUor object.
The historical $BVRI$ light curves of V582 Aur from all available photometric observations are plotted in Fig. 2. 
On the figure, the filled diamonds represent our CCD observations from Rozhen end Skinakas observatories,
the filled circles photographic data from the Asiago Schmidt telescope,
the filled triangles photographic data from the Kiso Schmidt telescope,
and the filled squares photographic data from the Palomar Schmidt telescope.

\begin{figure}[]
\begin{center}
 \includegraphics[width=13cm]{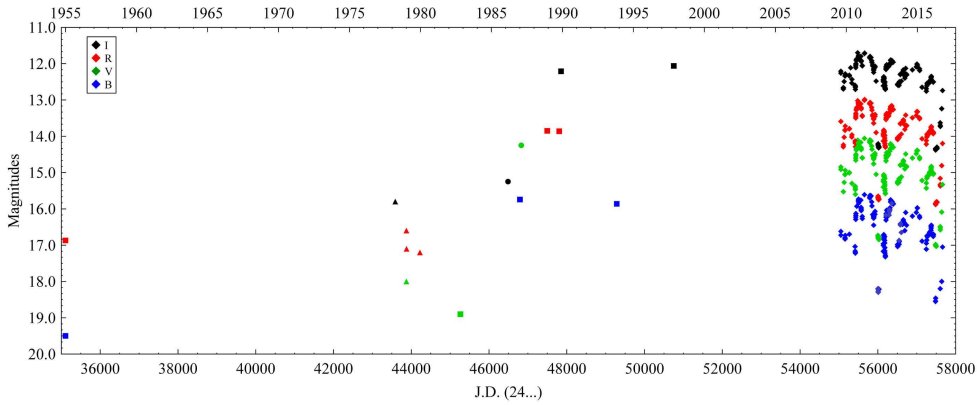} 
 \caption{Historical $BVRI$ light curves of V582 Aur for the period December 1954 $-$ October 2016}
   \label{fig2}
\end{center}
\end{figure}

The results of lasted for six years photometric monitoring of V582 Mon show extremely strong variability that is not seen in other FUor objects.
We suggest that the strong photometric variability can be explained by 1) time-variable extinction or 2) changes in accretion rate from the circumstellar disk onto the stellar surface. 
During the large drops in brightness, an appearance of dust particles in the immediate circumstellar environment of the star and change of the shape of basic spectral lines from absorption to emission was registered (\cite[Semkov et al. 2013]{Semkov2013}).

\subsection{V900 Mon}

The variability of V582 Aur was discovered by the amateur astronomer Jim Thommes.
Based on detailed multi-wavelength study of the star \cite[Reipurth et al. (2012)]{Reipurth2012} reach the conclusion that V582 Aur belongs to the group of FUor objects.
According to the authors the outburst of the star occurred between 1953 and 2009.
Recently \cite[Varricatt et al. (2015)]{Varricatt2015} registered a rise in the brightness of V900 Mon in the infrared.

\begin{figure}[]
\begin{center}
 \includegraphics[width=13cm]{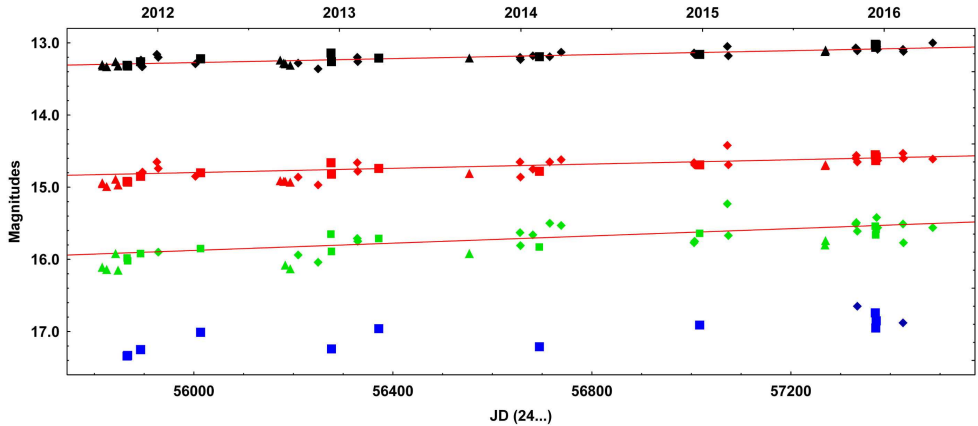} 
 \caption{$BVRI$ light curves of V900 Mon for the period August 2011 $-$ March 2016}
   \label{fig3}
\end{center}
\end{figure}

Our photometric monitoring of V900 Mon during the period from 2011 to 2016 shows a gradual increase in the brightness (Fig. 3).
Our search in the Digitized Sky Surveys shows that the star was registered at minimum light on the photographic plates obtained on 8 Jan. 1989 ($R$) and 10 Feb. 1985 ($I$). 
Hence the rise in the brightness of V900 Mon began after 1985 and continued in the recent years.

\end{document}